\documentclass[12pt]{article}

\def\beqq{\begin{equation}}   \def\eeqq{\end{equation}}
 \def\bea{\begin{eqnarray}}   \def\eea{\end{eqnarray}}
 
\newcommand{\MeV}{\,\mbox{MeV}}

\begin{document}
\begin{flushright} 
UND-HEP-04 -BIG\hspace*{.08em}09\\
hep-ph/0412041\\
\end{flushright}

\vspace*{4mm}

\centerline{\large\bf "I Have Come to Praise Charm, not Bury it"
\footnote{With due apologies to the `Bard'}
\footnote{Invited talk given at FPCP04 in Daegu (Korea), Oct. 5 - 10, 2004}} 

\vskip 0.3cm \centerline{I. I. Bigi} 

\vskip 0.3cm 
\centerline{Dept. of Physics, University of Notre Dame du Lac, Notre Dame, IN 46556, U.S.A.}
\centerline{e-mail: ibigi@nd.edu}
\vskip 1.0cm

\centerline{\bf Abstract}
The case for further dedicated studies of charm dynamics is presented as driven by three  
complementary motivations: core lessons on QCD and nonperturbative dynamics in general can be learnt; those can be applied with great profit in analyses of $B$ decays; charm decays offer a novel portal to New Physics in particular through CP studies.

\section{Introduction}
\label{INTRO}

There is the feeling in the HEP community that while the study of charm physics had a glorious 
past -- it drove the paradigm shift towards seeing quarks as real dynamical entities rather than objects 
of mathematical convenience thus providing essential support for the acceptance of the Standard Model (SM)  -- it has no future with its dull electroweak SM 
phenomenology: its CKM parameters are known, $D^0 - \bar D^0$ oscillations slow at best, CP asymmetries small and loop driven decays extremely rare. 
Yet I will stress that {\em dedicated} charm studies are full of promise due to a triple motivation: 
{\bf (i)} 
They will provide novel insights into the nonperturbative dynamics of QCD. 
{\bf (ii)} 
They will calibrate the theoretical tools for treating $B$ decays. 
{\bf (iii)} 
Certain charm transitions open a novel window onto New Dynamics. 

The accuracy of the theoretical description is of essential importance in these endeavours. After sketching the theoretical tools and listing lessons to be learnt on QCD in Sect.\ref{THEORY}, I address searches for New Physics in Sect.\ref{NEW} including a short comment on 
$\tau$ decays before concluding. I can touch on the various issues only briefly. Much 
more comprehensive discussions and references can be found in Ref.\cite{CICERONE}.

\section{Theory and its Tools}
\label{THEORY}

While we do not have a theory 
{\em of} charm -- i.e. why charm is the way it is -- we do have several theoretical tools {\em for} charm -- i.e. for treating charm dynamics. Its mass scale puts it somewhere between the worlds of bona fide heavy and light flavours. The accumulated evidence is that charm is `somewhat' heavy as naively expected. Quark models are still a useful tool for training our intuition and diagnosing results from lattice QCD (LQCD), but not reliable enough for final answers. 
Heavy quark expansions (HQE) based on expansions in powers of $1/m_c$ are a priori suspect, 
since the charm quark mass exceeds the scale of nonperturbative dynamics only moderately. 
Yet HQE work quite well for 
{\em inclusive} transitions like lifetimes -- unlike light cone sum rules for 
{\em exclusive} semileptonic decays, which fail.  
This could be due to the fact that the leading nonperturbative contributions to the former start in order $1/m_c^2$ involving local operators only, while the latter contain ${\cal O}(1/m_c)$ terms and  
{\em non}local correlators. 

LQCD is the only existing framework holding out the promise for a truly quantitative 
treatment of charm {\em hadrons} that can be improved {\em systematically} \cite{OKA}. Furthermore 
only LQCD can approach the charm scale both from below and above; hopefully charm 
will emerge as a firm `bridge' between the treatment of heavy and light flavours. At the same time the 
unchecked monopoly of a single theoretical technology to deal with nonperturbative dynamics should 
be viewed by its consumers with serious apprehension despite the siren songs of its producers. Its is therefore essential that the predictions of LQCD be subjected to a whole battery of precise experimental tests, and actually to a whole battery of the. This is happening for charm dynamics due to the comprehensive high quality data being obtained by  CLEO-c \cite{MILLER} to be joined by BES III later on and the beauty factories BELLE \& BABAR. 

\subsection{Lessons on QCD}
\label{QCD}

It is no longer adequate to talk about the mass of the charm quark {\em per se} and identify it with the parameter that appears in a quark {\em model}. A clean definition that can pass muster by field theory has to be given. The $\overline{MS}$ mass satisfies this criterion, and one finds 
\beqq 
\bar m_c(m_c) = \left\{
\begin{array}{ll} 1.19\pm 0.11\; {\rm GeV} & {\rm Ref.\cite{EIDE}}\\
1.14 \pm 0.1 \; {\rm GeV} & {\rm Ref.\cite{OLIVER}}
\end{array}
\right. \;  ; 
\label{MC}
\eeqq 
the first value is based on charmonium sum rules and the second one on moments of semileptonic $B$ decays. The fact that the numbers coming from systematically different observables agree so well supports the a priori conjecture that charm quarks can be treated as `somewhat heavy', since the scale of nonperturbative dynamics can be characterized by 
$\mu _{had} \sim 700 {\rm MeV} \sim N_C \Lambda _{QCD}$. 
Another indirect one is that about two thirds of $\Gamma (B \to l \nu X_c)$ is given by the two final states 
$D$ \& $D^*$, which are the ground states in the classification of Heavy Quark Symmetry. 

\subsubsection{Inclusive Rates}
\label{INCL}

The measured lifetimes of the seven $C=1$ charm hadrons provide a more quantitative measure for the heaviness of charm.  While a priori the HQE might be expected to fail even on the semiquantitative level since $\mu_{had}/m_c \sim 1/2$ (see Eq.(\ref{MC})), it works surprisingly well in describing the lifetime ratios even for baryons 
(see Table \ref{TABLECHARM}), 
except for $\tau (\Xi_c^+)$ being about 50 \% longer than predicted. 
\begin{table}
\begin{tabular} {|l|l|l|l|}
\hline
 & $1/m_c$ expect. & theory comments & data \\  
\hline  
\hline  
$\frac{\tau (D^+)}{\tau (D^0)}$ &  
$\sim 1+\left( \frac{f_D}{200\; \MeV} \right)^2 \sim 2.4$  
& PI dominant               & $2.54 \pm 0.01$  \\
\hline  
$\frac{\tau (D_s^+)}{\tau (D^0)}$ & 0.9 - 1.3[1.0 - 1.07] & 
{\em with} [{\em without}] WA  & $1.22 \pm 0.02$ \\
\hline  
$\frac{\tau (\Lambda _c^+)}{\tau (D^0)}$ & $\sim 0.5$  
& quark model matrix elements          & $0.49 \pm 0.01$ \\  
\hline  
$\frac{\tau (\Xi _c^+)}{\tau (\Lambda _c^+)}$ & $\sim 1.3 - 1.7$ &  
ditto                                  &  $2.2 \pm 0.1$\\
\hline  
$\frac{\tau (\Lambda _c^+)}{\tau (\Xi _c^0)}$ & $\sim 1.6 - 2.2$ &  
ditto                                  & $2.0 \pm 0.4$ \\
\hline
$\frac{\tau (\Xi _c^+)}{\tau (\Xi _c^0)}$ & $\sim 2.8$ &  
ditto                                  & $4.5 \pm 0.9$ \\
\hline  
$\frac{\tau (\Xi _c^+)}{\tau (\Omega _c)}$ & $\sim 4$ &  
ditto                                  & $5.8 \pm 0.9$ \\
\hline  
$\frac{\tau (\Xi _c^0)}{\tau (\Omega _c)}$ & $\sim 1.4$ &  
ditto                                  & $1.42 \pm 0.14$ \\
\hline  
\end{tabular}
\centering
\caption{Lifetime ratios of charm hadrons \cite{CICERONE}}  
\label{TABLECHARM}  
\end{table}
This agreement should be viewed as quite nontrivial, since these lifetimes span more than an order 
of magnitude between the shortest and longest: $\tau (D^+)/\tau (\Omega_c) \simeq 14$. 

The SELEX collab. has reported candidates for weakly decaying double charm baryons 
\cite{SELEX}. It is my judgment that those candidates cannot be $C=2$ baryons since their reported 
lifetimes are too short and do not show the expected hierarchy \cite{CICERONE}. 

$B_c$ mesons live in the worlds of beauty as well as of charm. While it had been suggested that 
binding energy effects lead to a `long' lifetime above 1 $psec$, the HQE predicts a `short' 
one: $\tau (B_c) \sim 0.3 - 0.7$ $psec$ with -- unlike in life -- charm fading faster than beauty \cite{MICHEL}. This prediction was supported by a 
first measurement by CDF and has been confirmed by D0 \cite{NOM}: 
\beqq 
\tau (B_c) = 0.45 ^{+0.12}_{-0.10} \pm 0.12 \; psec
\eeqq 
Another 
nontrivial HQE prediction is that the full semileptonic widths  of charm {\em baryons} are 
{\em far from universal}  -- unlike for charm {\em mesons}. The semileptonic branching ratios of baryons thus do not reflect their lifetimes. It would be highly desirable 
to measure BR$_{SL}(\Lambda_c)$ and BR$_{SL}(\Xi_c^{0,+})$. 

While $\Gamma _{SL}(D)$ is ill-suited to determine $|V(cs)|$ precisely, it is an interesting challenge to infer $|V(cd)/V(cs)|$ from the shape of inclusive lepton spectra in $D^0/D^+/D_s^+\to l \nu X_{s,d}$. 

\subsubsection{Exclusive Channels}
\label{EXCL}
 
The widths for $D^+/D_s^+ \to l^+ \nu$ with $l=\mu, \; \tau$ are controlled by the decay constants 
$f_D$ and $f_{D_s}$ leading to the following predictions: 
\bea 
{\rm BR}(D^+ \to \tau ^+ \nu) &\simeq& 1.1 \cdot 10^{-3} \left(\frac{f_D}{220\; \MeV}\right) ^2 \\
{\rm BR}(D^+ \to \mu ^+ \nu) &\simeq& 4.3 \cdot 10^{-4} \left(\frac{f_D}{220\; \MeV}\right) ^2  \\
{\rm BR}(D_s^+ \to \tau ^+ \nu) &\simeq& 5.1 \cdot 10^{-2} \left(\frac{f_{D_s}}{250\; \MeV}\right) ^2 \\
{\rm BR}(D_s^+ \to \mu ^+ \nu) &\simeq& 5.2 \cdot 10^{-3} \left(\frac{f_{D_s}}{250\; \MeV}\right) ^2
\eea
The CLEO-c collab. expects to measure these rates and thus $f_{D,D_s}$ with an uncertainty 
not exceeding a very few percent and compare it with future LQCD predictions of commensurate 
quality. Both of these goals look quite attainable and would open the era of {\em precision} tests of our 
understanding of nonperturbative dynamics. 

As far as other {\em exclusive} decays are concerned, theoretical tools exist only for semileptonic 
[nonleptonic] modes with one [two] hadron[s]/resonance[s] in the final states. Since the amplitudes for 
$D\to l \nu K[\pi]$ etc. depend on $|V(cs)[V(cd)] f_+^{K[\pi]}(q^2)$, there is a dual motivation 
for a careful analysis: 
accepting the values of $V(cs)$ and $V(cd)$ inferred from other processes or 
from three-family unitarity one extracts the formfactor, which can then be compared in its normalization as well as $q^2$ dependence with LQCD results; or one can employ the latter's prediction to infer the size of $V(cs)$ and $V(cd)$. For that purpose the level of accuracy has to be high to make it competitive. The theoretical prediction for the formfactor can of course be cross checked through its $q^2$ dependence. Yet that require very precise data since the range in 
$q^2$ is quite limited. It will be essential to do such an analysis for $D^0$, $D^+$ and  
$D_s^+$ Cabibbo allowed as well as suppressed modes and find consistent values for $V(cs)$ and $V(cd)$ before they can be accepted. The CLEO-c program is well equipped to perform such studies 
\cite{MILLER}.  Measuring $D^+/D_s^+ \to l \nu \eta/\eta^{\prime}$ can give us novel information  of the wavefunctions 
of $\eta$ and $\eta^{\prime}$; one can also search for glueball candidates $G$ in 
$D^+/D_s^+ \to l \nu G$.

The treatment of two-body nonleptonic decays poses a formidable theoretical challenge. It would make hardly any sense to rely on {\em pQCD}; the framework of {\em QCD factorization} should be tried, although it might  
fail due to its ${\cal O}(1/m_c)$ contributions, which are beyond theoretical control. The pioneering Blok-Shifman analysis \cite{BLOK} based on QCD sum rules should be updated and refined by including $SU(3)_{Fl}$ breaking. A meaningful LQCD analysis has to be fully unquenched. In conclusion: the only tools available at present are quark models; yet their findings have to be taken with quite a rock of salt. For a description of nonleptonic charm decays to claim reliability, it has to succeed on the Cabibbo allowed as well as singly or doubly Cabibbo suppressed levels, including resonant final states with more than one neutral hadron.  

Establishing theoretical control over QCD's dynamics will teach us important lessons about nonperturbative dynamics in general, as is relevant for New Physics models based on technicolour to cite but one example.

\subsection{`Tooling up' for $B$ Studies}
\label{BSTUD}

The nonperturbative dynamics driving the exclusive transitions 
$B\to l \nu D/D^*$ is characterized by the scale $m_c$, not $m_b$. Studying charm decays can thus 
provide important, at times even essential lessons on $B$ decays. A few examples 
of this connection have to suffice. 

Once LQCD's predictions for $f_{D,D_s}$ have (hopefully) been validated on the very few percent accuracy level, one can scale them up with confidence for $f_{B,B_s}$ with great phenomenological 
benefit for our theoretical interpretation of $B_d - \bar B_d$ and $B_s - \bar B_s$ oscillations. 

An analogous strategy will be pursued for exclusive semileptonic $B$ decays. Once LQCD's 
results on the form factors for $D/D_s \to l \nu M$, $M= K^*, K, \rho, \pi, \eta, ...$ have been validated 
in their normalization as well as their $q^2$ dependence, one can extend these methods with enhanced confidence to the treatment of $B \to l \nu \pi/\rho$ etc. to extract $|V(ub)|$.

\subsubsection{The `3/2  $<$ 1/2' Puzzle}
\label{PUZZLE}

While the lessons sketched above are common knowledge, this one is not. Semileptonic 
$B$ decays present us with three motivations to gain a better understanding of charm 
spectroscopy, the first two experimental and the third one experimental as well as theoretical: (i) Extracting $\Gamma (B \to l \nu X_c)$ and its errors from the data with high 
accuracy requires good understanding of possible final states to determine detector efficiencies etc.;  
(ii) likewise for $B\to l \nu D/D^*$ and the feed down  from higher charm resonances. 
(iii) There are classes of sum rules derived from QCD proper that  relate the 
heavy quark parameters appearing in the OPE for inclusive $B\to l \nu X_c$ -- like 
$\mu_{\pi}^2$, $\mu _G^2$ etc. -- with restricted sums over exclusive channels. They 
provide rigorous definitions, inequalities and experimental constraints \cite{HQSR};  
e.g.:
\bea 
\frac{1}{2} &=& 2 \sum _m |\tau _{3/2}(m)|^2 - 2 \sum _n |\tau _{1/2}(n)|^2  \\
\mu_G^2(\mu) &=& 2 \sum _m \epsilon _m^2|\tau _{3/2}(m)|^2 - 2 \sum _n 
\epsilon _n^2|\tau _{1/2}(n)|^2 \; , 
\label{SUMRULES}
\eea
where $\tau_{1/2}$,  $\tau _{3/2}$ are the amplitudes for 
$B\to l \nu D(s_q)$ with $D(s_q)$ a hadronic system beyond the $D$ and $D^*$,  
$s_q = 1/2 \& 3/2$ the angular momentum carried by the light degrees of freedom in 
$D(s_q)$ and $\epsilon_m$ the excitation energy of the $m$th such system above the $D$  
with $\epsilon _m \leq \mu$. Eq.(\ref{SUMRULES}) manifestly shows that the 
$s_q = 3/2$ contributions have to dominate over the $s_q = 1/2$ ones. There were indications in early 
data -- mainly from nonleptonic decays treated under the assumption of factorization -- that this hierarchy is not obeyed by the lowest $P$ wave states of which there are four: two 
narrow $3/2$ ($D_1,D_2^*$) and two broad $1/2$ states ($D_0^*,D_1^{\prime}$) \cite{MINNE}. It would be conceivable 
mathematically that higher resonances would rectify the situation, yet that seems a very contrived 
solution. Recent BELLE data \cite{BELLE32} seem to be consistent with the sum rules. It is important to obtain conclusive data on this issue in semileptonic $B$ decays 
\cite{NOM}. 

Understanding this spectroscopy is important not only in its own right and because heavy quark theory makes nontrivial predictions on it. As stated above it is needed to have full control 
over the measurements of $B\to l \nu X_c$ (and its moments) as well as $B\to l \nu D^*$.  This is desirable also for an exotic scenario: finding the values for $|V(cb)|$ as extracted from 
$B\to l \nu X_c$ and $B\to l \nu D^*$ to disagree -- or the measured moments of the former 
{\em not} be described by the same set of heavy quark parameters --, might not point to a true failure of the theoretical description. Instead it might signal the presence of {\em right}-handed charged current 
couplings for the $b$ quark!

\section{Searching for New Physics}
\label{NEW}

It has often been said that with the `dull' SM weak phenomenology 
for charm -- slow $D^0 - \bar D^0$ oscillations, small CP asymmetries -- 
charm studies allow almost `zero-background' searches 
for New Physics. Yet this statement has to be updated carefully since experiments over the last ten years have bounded the oscillation parameters $x_D$, $y_D$ to fall below  very few \% and direct CP asymmetries below several \%. One should take note that charm is the only {\em up-}type quark allowing the full range of 
probes for New Physics, including flavour changing neutral currents: while top quarks do not hadronize \cite{RAPALLO}, in the $u$ quark sector 
you cannot have $\pi^0 - \pi^0$ oscillations and many CP asymmetries are already ruled out by 
CPT invariance. My basic contention  is the following: {\em Charm transitions are a unique 
portal for obtaining a novel access to flavour dynamics with the experimental situation 
being a priori favourable (except for the lack of Cabibbo suppression)!}

\subsection{$D^0 - \bar D^0$ Oscillations}
\label{DOSC}

$D^0 - \bar D^0$ oscillations represent a subtle quantum mechanical phenomenon of great 
practical importance: it can have a significant impact on extracting the CKM phase 
$\phi_1/\gamma$ from $B^{\pm} \to D^{neut}K^{\pm}$; it provides a probe for New Physics, albeit an ambiguous one; it represents an important ingredient for CP asymmetries arising in $D^0$ decays due to New Physics. 

These phenomena can be characterized by two quantities, namely 
$x_D = \frac{\Delta M_D}{\Gamma_D}$ and $y_D =\frac{\Delta \Gamma_D}{2\Gamma_D}$.  
Oscillations  are slowed down in the SM due to GIM suppression and $SU(3)_{fl}$ symmetry. 
Comparing a {\em conservative} SM bound with the present data  
\beqq 
x_D(SM), y_D(SM) < {\cal O}(0.01)  \; \; vs. \; \; 
\left. x_D\right|_{exp}  < 0.03 \; , \; \;  \left. y_D\right|_{exp} = 0.01 \pm 0.005 
\label{DOSC}
\eeqq 
we conclude that the search has just now begun. There exists a considerable literature -- yet 
typically with several ad-hoc assumptions concerning the nonperturbative dynamics. It is widely understood that the usual quark box diagram is utterly irrelevant due to its untypically severe 
GIM suppression $(m_s/m_c)^4$. 
A systematic 
analysis based on an OPE has been given in Ref.\cite{BUDOSC} in terms of powers of 
$1/m_c$ and $m_s$. Contributions from higher-dimensional operators with a much softer 
GIM reduction of $(m_s/\mu_{had})^2$ (even $m_s/\mu_{had}$ terms could arise) due to `condensate'  terms in the OPE  yield 
\beqq 
\left. x_D (SM)\right|_{OPE}, \; \left. y_D (SM)\right|_{OPE} \sim {\cal O}(10^{-3}) \; . 
\eeqq 
Ref.\cite{FALK} finds very similar numbers, albeit in a quite different approach. 

While one predicts similar numbers for $x_D(SM)$ and $y_D(SM)$, one should keep in mind 
that they arise in very different dynamical environments. $\Delta M_D$ is generated from off-shell intermediate states and thus is sensitive to New Physics, which could produce 
$x_D \sim {\cal O}(10^{-2})$. $\Delta \Gamma_D$ on the other hand is shaped by on-shell intermediate 
states; while it is hardly sensitive to New Physics, it involves much less averaging or `smearing' than 
$\Delta M_D$ making it thus much more vulnerable to violations of quark-hadron duality. Observing 
$y_D \sim 10^{-3}$ together with $x_D \sim 0.01$ would provide intriguing, though not conclusive 
evidence for New Physics, while $y_D \sim 0.01 \sim x_D$ would pose a true conundrum for its 
interpretation. Yet even those have to be measured for a proper analysis of $B^{\pm} \to D^{neut}K^{\pm}$, preferably down to the $10^{-3}$ level. 

\subsection{CP Violation}
\label{CPV}

Since the baryon number of the Universe implies the existence of New Physics in CP violating dynamics, it would be unwise not to undertake dedicated searches for CP asymmetries in 
charm decays, where the `background' from known physics is small: within the SM the effective weak phase is highly diluted, namely $\sim {\cal O}(\lambda ^4)$, and it can 
arise only in singly Cabibbo suppressed transitions, where one  
expects them to reach the 0.1 \% level; significantly larger values would signal New Physics.  
{\em Any} asymmetry in Cabibbo 
allowed or doubly suppressed channels requires the intervention of New Physics -- except for 
$D^{\pm}\to K_S\pi ^{\pm}$ \cite{CICERONE}, where the CP impurity in $K_S$ induces an asymmetry of 
$3.3\cdot 10^{-3}$. Several facts actually favour such searches: strong phase shifts 
required for direct CP violation to emerge in partial widths are in general large as are the branching ratios into relevant modes;  
finally CP asymmetries can be linear in New Physics amplitudes thus enhancing sensitivity to the 
latter.  As said above, the benchmark scale for KM asymmetries in singly Cabibbo suppressed 
partial widths is 
$0.1\%$. This does not exclude the possibility that CKM dynamics might exceptionally generate an \
asymmetry as `large' as 1\% in some special cases. It is therefore essential to analyze a host of 
channels. 

Decays to final states of {\em more than} two pseudoscalar or one pseudoscalar and one vector meson contain 
more dynamical information than given by their  widths; their distributions as described by Dalitz plots 
or {\em T-odd} moments can exhibit CP asymmetries that can be considerably larger than those for the 
width. Final state interactions while not necessary for the emergence of such effects, can fake a signal; 
yet that can be disentangled by comparing {\em T-odd} moments for CP conjugate modes. I view this as a very promising avenue, where we still have to develop the most effective analysis tools for small 
asymmetries.

CP violation involving $D^0 - \bar D^0$ oscillations can be searched for in final states common to $D^0$ 
and $\bar D^0$ decays like CP eigenstates -- $D^0 \to K_S\phi$, $K^+K^-$, $\pi^+\pi^-$ -- or 
doubly Cabibbo suppressed modes -- $D^0 \to K^+\pi^-$. The CP asymmetry is controlled by  
sin$\Delta m_Dt$ $\cdot$ Im$(q/p)\bar \rho (D\to f)$; within the SM both factors are small, namely 
$\sim {\cal O}(10^{-3})$, making such an asymmetry unobservably tiny -- unless there is New Physics! 
One should note 
that this observable is linear in $x_D$ rather than quadratic as for CP insensitive quantities.  
$D^0 - \bar D^0$ oscillations, CP violation and New Physics might thus be discovered simultaneously in a transition. 

One wants to reach the level at which SM effects are 
likely to emerge, namely down to time-{\em dependent} CP asymmetries 
in $D^0 \to K_S\phi$, $K^+K^-$, $\pi^+\pi^-$ [$K^+\pi^-$] down to $10^{-5}$ [$10^{-4}$] and {\em direct} 
CP asymmetries in partial widths and Dalitz plots down to $10^{-3}$.

\subsection{Short Comment on $\tau$ Decays}
\label{TAU}

Like charm hadrons the $\tau $ lepton is often viewed as  a system with a great past, but hardly a 
future. Again I think this is a very misguided view and I will illustrate it with two examples. 

If baryogenesis is a secondary phenomenon driven by primary leptogenesis, one needs CP violation 
in the lepton sector. In my judgment $\tau$ decays -- together with electric dipole moments for leptons and possibly $\nu$ oscillations -- provide the best stage to search for other manifestations of these dynamics directly. 

Searching for $\tau ^{\pm} \to \mu ^{\pm} \mu ^+\mu ^-$ (and its variants) -- 
processes forbidden in the SM -- is particularly intriguing, since it involves only `down-type' leptons 
of the second and third family and is thus the complete analogy of the quark lepton process 
$b \to s \bar s s$ driving $B_s \to \phi K_S$, which has recently attracted such strong attention. 
Following this analogy literally one guestimates ${\rm BR}(\tau \to 3 \mu) \sim 10^{-8}$ to be 
compared with the present bound from BELLE \cite{BELLETAU} 
\beqq 
{\rm BR}(\tau \to 3 \mu) \leq 2\cdot 10^{-7} \; . 
\eeqq
It would be very interesting to know what the 
$\tau$ production rate at the hadronic colliders is and whether they could be competitive or even superior with the $B$ factories in such a search.

\section{Conclusions and Outlook}
\label{CON}

The SM has scored qualitatively new successes in flavour dynamics since the beginning of this millenium \cite{BIGICKM}. Yet we have to admit that we `know so much, yet understand so little'; i.e., the SM provides an incomplete picture of Nature's 
`Grand Design'. I firmly believe that we need further hints from Nature to get a more complete picture. 
Dedicated and comprehensive studies  of the decays of charm mesons and $\tau$ leptons will prove 
essential in our endeavour. They will sharpen our understanding of QCD and nonperturbative 
dynamics in general and validate our tools for treating $B$ decays with the required accuracy. 
Last, not least they offer the persistent student the promise to identify the intervention of New Physics, 
in particular in the area of CP violation. The observed baryon number of the universe and the `strong CP problem' provide intriguing evidence for the presence of New Physics.

\section*{Acknowledgments}
I would like to thank the organizers of FPCP04 for creating such a fine meeting. 
This work was supported by the NSF under grant numbers PHY00-87419 \& PHY03-55098.


%
\label{IBigiEnd}

\end{document}